\documentclass[sigplan,nonacm]{acmart}

\AtBeginDocument{%
  }

\setcopyright{acmlicensed}
\copyrightyear{2018}
\acmYear{2018}
\acmDOI{XXXXXXX.XXXXXXX}
\acmConference[Conference acronym 'XX]{Make sure to enter the correct
  conference title from your rights confirmation email}{June 03--05,
  2018}{Woodstock, NY}
\acmISBN{978-1-4503-XXXX-X/2018/06}

\usepackage{makecell}
\usepackage{listings}
\usepackage{multirow}
\usepackage{mdframed}
\usepackage{xcolor}
\usepackage{siunitx}
\usepackage{url}
\definecolor{grey}{RGB}{128,128,128}

\newcounter{observation}
\newenvironment{observation}{%
    \refstepcounter{observation}%
    \begin{mdframed}[linecolor=black,linewidth=1pt,backgroundcolor=gray!10,skipabove=\baselineskip,skipbelow=\baselineskip]%
    \small
    \textbf{Observation \theobservation:}~\itshape}
    {\end{mdframed}}

\begin{document}


\title{Reliability Analysis of Fully Homomorphic Encryption Systems Under Memory Faults}

\author{Rian Adam Rajagede}
\affiliation{%
  \institution{University of Central Florida}
  \city{Orlando}
  \state{FL}
  \country{USA}
}
\email{rian@ucf.edu}

\author{Yan Solihin}
\affiliation{%
  \institution{University of Central Florida}
  \city{Orlando}
  \state{FL}
  \country{USA}
}
\email{yan.solihin@ucf.edu}









\begin{abstract}
Fully Homomorphic Encryption (FHE) represents a paradigm shift in cryptography, enabling computation directly on encrypted data and unlocking privacy-critical computation. Despite being increasingly deployed in real platforms, the reliability aspects of FHE systems, especially how they respond to faults, have been mostly neglected. This paper aims to better understand of how FHE computation behaves in the presence of memory faults, both in terms of individual operations as well as at the level of applications, for different FHE schemes. Finally, we investigate how effective traditional and FHE-specific fault mitigation techniques are. 
\end{abstract}




\maketitle

\section{Introduction}
Fully Homomorphic Encryption (FHE) is a revolutionary cryptographic technique that allows computation on encrypted data without requiring decryption first~\cite{gentry2009fully, cheon2017homomorphic, brakerski2014leveled}. This capability makes FHE very attractive for privacy-sensitive applications like healthcare analytics, financial services, and secure cloud computing \cite{hong2025secure,han2019logistic,wood2020homomorphic}, where protecting data confidentiality is critical. Unlike traditional encryption methods that require data to be decrypted before processing, FHE enables third parties to perform complex operations directly on encrypted data, producing encrypted results that only the data owner can decrypt. This fundamental advancement in cryptography promises to transform how sensitive information is processed. Furthermore, groundbreaking progress has been made to address its high execution time. For example, over the past two years, the latency for deep neural network inference on ResNet-20 has decreased from 88,320 seconds to just 15.9 seconds~\cite{fan2023tensorfhe, samardzic2022craterlake,kim2022ark}. As a result, FHE is increasingly deployed in the real world, including products from Duality~\cite{openfhe},  Microsoft~\cite{sealcrypto}, and Apple~\cite{apple, asi2024scalableFHEAPPLE}. 

However, despite being increasingly deployed in real platforms, the reliability aspects of FHE systems, in particular how they respond to memory faults, have been mostly neglected in the research community. Memory faults typically manifest themselves as "random" bitflips, and their likelihood is affected by technology scaling, cosmic radiation, aging hardware, temperature variations, and voltage~\cite{hwang-asplos12-CosmicRay, electric_bitflip, baumann2005radiation, kim2014flipping}. The effects of memory faults on FHE computation can be severe. FHE only guarantees privacy but not integrity, hence, a memory fault may lead to a silent data corruption (SDC) where users receive incorrect computation results unknowingly. Furthermore, FHE is especially vulnerable to memory faults, because it slows down both execution~\cite{zhang2024sok,gouert2023sok,yudha2024boostcom, jung2021over} and expands the data size, both by several orders of magnitude~\cite{gentry2012homomorphic, canteaut2018stream, cheon2019full}. Consequently, it is substantially more likely to experience random memory bit flips during the course of their execution, both {\em temporally} (due to the lengthened execution time) and {\em spatially} (due to the larger memory footprint). 

The goal of this paper is to get a better understanding of how FHE computation behaves in the presence of memory faults. There are several reasons why answering this question is interesting and important for providing clarity. First, for certain FHE schemes, one byte of plaintext data expands to roughly 1 million bits of ciphertext. A single bit flip in a byte of plaintext affects 12.5\% of the data, but it only affects 0.0001\% of the data in the ciphertext form. Hence, it is plausible that FHE computation is more fault-tolerant than plaintext computation. However, at the same time, FHE computation relies on complex mathematical structures, such as high-degree polynomial rings and large-modulus arithmetic, which may react to memory faults in nonlinear ways, potentially propagating a single bit flip into broader ciphertext corruption. Hence, is it possible that FHE computation may instead be less fault-tolerant? Second, FHE may be used for private machine learning (ML). Neural network-based ML exhibits resilience to random weight perturbations~\cite{rakin2019bit, agarwal2023resilience, DNN_bitflip_2, mukherjee2003systematic, rajagede2025naper}. Is this natural fault tolerance still the case with ML on FHE?  Third, FHE has different schemes that differ on the native data types, e.g., BGV (integers) vs. CKKS (floating points). Do different FHE schemes exhibit different behavior under memory faults? Fourth, among known techniques for fault tolerance, including error correcting code (ECC), checksum, and modular redundancy, how effective are they in detecting or recovering from memory faults? What are their computation overheads?  What are their relative advantages and drawbacks? 

In this paper, we present the first comprehensive study of the fault tolerance behavior of  FHE computation under memory faults. We answer three fundamental research questions: (1) {\em How do individual FHE operations respond to memory faults?} We investigate how a memory fault affects individual FHE operations, such as a single addition, multiplication, and rotation, and they may involve all ciphertext operands or mixed ciphertext/plaintext operands. We examine this question for two FHE schemes (CKKS and BGV), as well as for their various parameter choices.

(2) {\em How does Machine Learning (ML) inference with FHE respond to memory faults?} Looking beyond a single operation, we investigate how a fault impacts the output accuracy of privacy-preserving ML applications, including linear regression and neural networks, to understand how errors propagate through ML pipelines.

(3) {\em How do mitigation strategies compare for ensuring the reliability of FHE systems?} We evaluate their relative efficacy, costs (both latency and storage overheads), which scheme it applies to, whether it involves client cooperation, and discuss their relative advantages and drawbacks.

The remainder of this paper is structured as follows. Section 2 discusses related work, Section 3 provides background on FHE schemes, Section 4 describes our fault model, Section 5 presents our characterization study design, Section 6 details our experimental methodology, Section 7 presents our results, and finally Section 8 concludes the paper. 

\section{Related Works}

\paragraph{General fault tolerance techniques}
For general computation, well-known techniques for fault tolerance include error-detecting codes such as parity and checksums, and error-correction codes (ECC)~\cite{hamming1950error}. The most commonly implemented ECC for main memory in servers is Single Error Correction Double Detection (SECDED) can detect and correct single-bit errors and detect double-bit errors, adding 8 parity bits for 64-bit data words (12.5\% overhead), requiring specialized hardware but incurring minimal computational overhead during normal operation. Beyond memory faults, modular redundancy~\cite{lyons1962use} offers stronger protection by running identical copies of a system; with triple modular redundancy (TMR), majority voting is used to select the correct outputs. These techniques have different fault coverage and costs; for example, TMR more than triples resource requirements but covers more types of faults. While their uses have been extensively evaluated in plaintext computation, this paper evaluates their usage, advantages, and drawbacks for protecting FHE computation. 

\paragraph{FHE computation fault tolerance techniques} 
Some existing works address different concerns. For instance, Chillotti et al.\cite{chillotti2016attacking} examined how deliberately injected faults can be exploited as attack vectors against FHE systems, in the context of adversarial attacks to extract sensitive information. Specific to FHE computation reliability, there have been recent works to add verifiability to FHE. Verifiability techniques can detect faults, but the goal is broader, i.e., to ensure the correct FHE computation is performed. They typically integrate additional cryptographic techniques to FHE, for example zero knowledge proof (ZKP)~\cite{viand2023verifiable, chatel2022verifiable}, running FHE computation in an attested secure enclave using Trusted Execution Environment (TEE)~\cite{natarajan2021chex, coppolino2020vise}, and using a cryptographically fortified algorithm-based fault tolerance (ABFT)~\cite{santriaji2024dataseal}. They are generally very slow, adding one or more orders of magnitude slowdown, except DataSeal~\cite{santriaji2024dataseal}, which incurs very low performance overheads for sufficiently large inputs. However, DataSeal is not applicable to the computation model that we assume (the private circuit discussed in Section~\ref{subsec:fault_model}). Another work, X-Cipher~\cite{caulfield2024x}, protects FHE system availability from entire-node server failures on a multiple storage server system via erasure codes. X-Cipher does not characterize how memory faults affect FHE systems, how they propagate, and how they should be detected.



None of the works above analyzes how FHE computation behaves under memory faults. In contrast, our work presents a comprehensive study on how transient memory faults affect FHE systems, from FHE operations to the applications, how they propagate, how they can be detected and corrected, and analyzes the benefits and drawbacks of fault-tolerant techniques when applied to FHE computation. 

\section{Background}
\label{sec:bkground}

\begin{figure}[tbp]
\centerline{\includegraphics[width=0.45\textwidth]{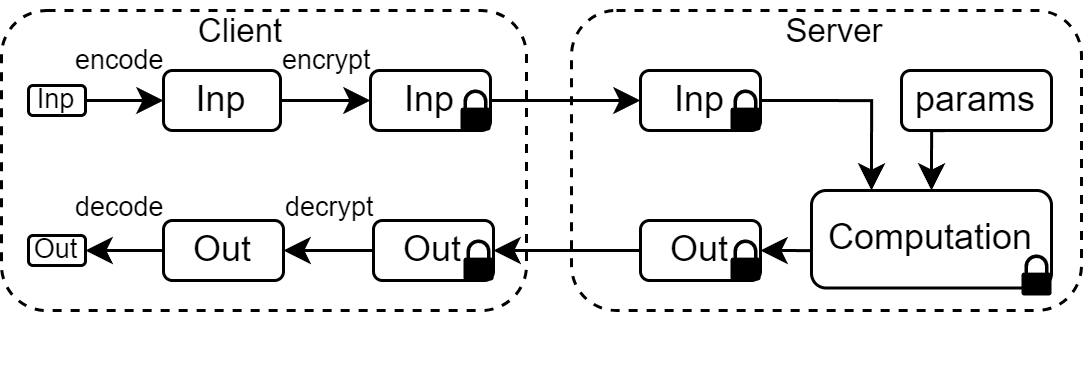}}
\caption{Illustration of FHE Systems}
\label{figsystem}
\end{figure}

\subsection{Fully Homomorphic Encryption}
Homomorphic encryption (HE) is a cryptographic technique where an operation performed in the ciphertext space yields an identical result as when the operation were performed in the plaintext space. HE allows a server to perform computation directly on ciphertext input data, yielding a ciphertext result that can only be decrypted by the client. If $m_1$ and $m_2$ denote plaintext data, $E$ the encryption function, while $\oplus_{c}$ and $\otimes_{p}$ denote an operation on the ciphertext and plaintext respectively, then HE guarantees that $E(m_1) \oplus_{c} E(m_2) = E(m_1 \oplus_{p} m_2)$. An FHE operation may be an addition, a multiplication, a rotation, etc. Fully Homomorphic Encryption (FHE) generally refers to a major advancement of homomorphic encryption that allows an arbitrarily long series of operations to be performed~\cite{gentry2009fully}. Operations performed with FHE are also referred to as {\em computation circuit}. 

FHE systems typically operate in a client-server model as illustrated in Figure~\ref{figsystem}. The client encodes then encrypts data before sending it as a message to the server, which performs computations directly on the encrypted data. In the server, parameters (also called "input" in~\cite{viand2023verifiable}) needed for performing computations (e.g., machine learning model weights) are usually stored as plaintext. Such "plaintext" is encoded in polynomial representations, and not in conventional non-FHE format. For example, with RNS, weights are encoded as an RNS representation matrix, which is about hundreds of times larger than the original message.

There are two FHE computation models from the client perspective: {\em public circuit}, where a client has a whitebox view of the computation provided by the server, and  {\em private circuit}~\cite{viand2023verifiable}, where a client has a blackbox view. The latter covers many service-oriented deployments, such as Machine Learning as a Service (MLaaS), where the service providers use proprietary models~\cite{wood2020homomorphic}.



\subsection{FHE Schemes}

FHE schemes primarily rely on the hardness of lattice problems, such as Ring-Learning With Errors (RLWE). Families of FHE schemes differ in their underlying plaintext structure, the types of operations they support, and their noise management techniques. Two of the most prominent schemes are: BGV~\cite{brakerski2014leveled} that supports integers and CKKS~\cite{cheon2017homomorphic} that operates natively on real and complex numbers.


\subsubsection{Mathematical Foundation}
Most FHE schemes use polynomial rings. A common choice is $R = \mathbb{Z}[X]/(\Phi_M(X))$, where $\Phi_M(X)$ is a cyclotomic polynomial, often $X^N+1$ (defining the ring dimension $N$, typically a power of 2). Plaintexts are polynomials with coefficients modulo the plaintext modulus $t$. Ciphertexts are typically pairs $(c_0, c_1)$ of polynomials with coefficients modulo a larger ciphertext modulus $q$.

\subsubsection{Noise Growth}
FHE schemes incorporate noise for security, which grows with each operation, especially multiplications. When noise exceeds a threshold, decryption fails.


\subsubsection{Encoding and Batching}
Raw data must be encoded into the plaintext ring $R_t$ before encryption. FHE schemes leverage the structure of cyclotomic rings to pack multiple data values into a single polynomial, enabling Single Instruction, Multiple Data (SIMD) operations. This is known as \textit{batching}. The encoding steps in BGV and CKKS are different based on how they handle different data types: 

{\em BGV Encoding:} Multiple integers are packed into 'slots' within a single plaintext polynomial $m \in R_t$ (using principles related to the Chinese Remainder Theorem). 

{\em CKKS Encoding:} Real or complex numbers are first scaled by a factor $\Delta$ (becoming approximate integers) and then packed into 'slots' of a polynomial $m$.

\subsubsection{Homomorphic Operations}
\label{heops}
FHE schemes support various operations directly on ciphertexts:

{\em Plaintext Addition (PADD):} Adding a plaintext polynomial $p$ to a ciphertext $\mathbf{ct}_1$:
    \begin{equation}
    \label{eq:padd}
    \mathbf{ct} + p = (c_{0} + p, c_{1}) \pmod{q}
    \end{equation}

{\em Plaintext Multiplication (PMUL):} Multiplying a ciphertext $\mathbf{ct}$ by a plaintext polynomial $p$:
    \begin{equation}
    \label{eq:pmul}
    \mathbf{ct} \cdot p = (c_{0} \cdot p, c_{1} \cdot p) \pmod{q}
    \end{equation}

{\em Ciphertext-Ciphertext Addition (HADD):} Given $\mathbf{ct}_1 = (c_{1,0}, c_{1,1})$ encrypting $m_1$ and $\mathbf{ct}_2 = (c_{2,0}, c_{2,1})$ encrypting $m_2$, their sum is computed component-wise:
    \begin{equation}
    \mathbf{ct}_1 + \mathbf{ct}_2 = (c_{1,0} + c_{2,0}, c_{1,1} + c_{2,1}) \pmod{q}
    \end{equation}

{\em Ciphertext-Ciphertext Multiplication (HMUL):} Their product initially forms a 3-component ciphertext:
   \begin{equation}
   \label{eq:hmul}
   \mathbf{ct}_{\text{mul}} = (c_{1,0}c_{2,0},c_{1,0}c_{2,1} + c_{1,1}c_{2,0}, c_{1,1}c_{2,1}) \pmod{q}
   \end{equation}
   Then this form is converted back to a standard 2-component ciphertext using \textit{relinearization} with a public relinearization key. In addition, to manage noise, BGV schemes typically perform \textit{modulus switching} (reducing the ciphertext modulus $q$). CKKS employs {\em rescaling} to correct the plaintext's scale (which changes from $\Delta$ to $\Delta^2$ upon multiplication).


{\em Homomorphic Rotation (HROT):} Permutes the values within the batched slots of a ciphertext. This is essential for operations requiring data movement across slots. It's achieved by applying a Galois automorphism to the ciphertext, which requires another special evaluation key. 

\subsection{Efficiency Techniques in FHE}
\label{sec:rns} 
Handling the very large integer coefficients of polynomials modulo $q$, as required for FHE security, is computationally demanding on standard hardware. Modern FHE libraries~\cite{sealcrypto, openfhe} employ two key optimization to manage this complexity. 

One such optimization is the use of {\em Residue Number System} (RNS). Instead of operating directly with the large modulus $q$, it is decomposed into a product of smaller, typically machine-word-sized, pairwise coprime primes: $q = q_0 q_1 \cdots q_L$. An integer $x \pmod q$ is then represented by its vector of residues $(x \pmod{q_0}, x \pmod{q_1}, \dots, x \pmod{q_L})$. Therefore, arithmetic operations like addition and multiplication modulo $q$ can be performed independently and in parallel on these smaller residues modulo each $q_i$. This RNS decomposition is applied coefficient-wise to the polynomials in the ring $R_q$.

Another optimization is the use of {\em Number Theoretic Transform} (NTT).  Polynomial multiplication is a fundamental operation in FHE but is computationally expensive in coefficient form (naive complexity $\mathcal{O}(N^2)$). The NTT, a specialized version of the Fast Fourier Transform (FFT) adapted for finite fields (specifically $\mathbb{Z}_{q_i}$), accelerates this significantly. It transforms a polynomial from its coefficient representation to an evaluation (or NTT) representation in $\mathcal{O}(N \log N)$ time. In the NTT domain, the multiplication of two polynomials becomes a simple element-wise multiplication. An inverse NTT converts the result back to coefficient form when needed.

These two techniques are typically combined. A polynomial $a(X) \in R_q$ is first decomposed into its RNS components $a_i(X) = a(X) \pmod{q_i}$ for $i=0, \dots, L$. Then, for efficient multiplication, each polynomial component $a_i(X)$ is often kept in its NTT representation. This results in a data structure for representing ciphertext polynomials that can be visualized as a matrix. Each row corresponds to an RNS modulus $q_i$, and the columns hold the coefficients of the NTT representation for that component polynomial. This structure, sometimes referred to as a Double-CRT representation~\cite{gentry2012homomorphic}.

\section{Fault Model}
\label{subsec:fault_model}

{\em FHE Computation Model:} We assume the FHE computation with a private circuit model, where the server provides services to the client without giving visibility to the client into the computational process performed by the server, including the circuit (algorithm) and the server input (e.g., model parameters). The client only receives the output after the server finishes all computations and may not necessarily be able to verify the computation correctness after decryption. On the other hand, the server cannot see the plaintext of the client data as the secret key is in the client.


{\em Fault Type and Scope:} We focus on faults affecting data stored in memory components, including cache and main memories (such as DRAM) in the client and server.  
Memory faults may be permanent (e.g. stuck-at, stuck-open faults), or transient (e.g. soft errors). We focus primarily on "random" memory bitflips, which may be caused by cosmic radiation, voltage fluctuations, thermal variations, or aging effects in memory cells. Our primary focus is server-side faults encompassing static data (e.g., DNN model parameters), intermediate outputs, and final outputs during FHE computation, but we also examine client-side faults affecting input data before encryption to understand how fault location impacts error propagation patterns. We assume that multi-bit error is a substantial possibility, as shown in a recent large-scale study of production servers of IBM BG/P and BG/L systems~\cite{hwang-asplos12-CosmicRay, jung2023predicting}. This paper focuses specifically on random, non-adversarial memory faults. We exclude deliberate fault injection attacks, adversarial fault patterns, or security-motivated fault injection scenarios from our study.

\section{Characterization Study Design}

\subsection{Analyzing the Reliability Behavior of FHE Computation}

\paragraph{Fault likelihood} FHE computation is different from regular plaintext computation in several ways. First, FHE computation is characterized by {\em large ciphertexts}. A single multiplication (depth $L=1$) of one byte of plaintext is encoded and encrypted into two large polynomials in a ring with dimension N=8192 (the lowest ring dimension to comply with the 128-bit security standard), with each coefficient typically the size of 8 bytes. This translates to 2 polynomials $\times$ 8192 coefficients $\times$ 2 moduli (for $L=1$) $\times$ 8 bytes = 262,144 bytes ciphertext for a single byte of plaintext, demonstrating a ciphertext expansion of about five orders of magnitude~\cite{gentry2012homomorphic, canteaut2018stream, cheon2019full}. Thus, the larger memory footprint exposes FHE to a higher probability of experiencing memory bit flips during the course of the computation. 


FHE computation is also {\em computationally expensive}; a simple operation can take 1000x to 10,000x longer than plaintext computation~\cite{yudha2024boostcom}. The much longer execution time also exposes FHE to a higher probability of experiencing memory bit flips. 

Taken together, the expanded memory footprint and execution time result in between 100 million to 1 billion times higher likelihood for the FHE computation to suffer from a memory fault compared to the equivalent plaintext computation, for an equivalent computation task. 

\paragraph{Fault propagation}
Besides increased exposure to memory faults, FHE computation employs {\em complex mathematical} structures with many multiple processing stages operating on large polynomials, even for a single operation. Not all data will be in the form of ciphertext, for example, ML model parameters may be in the form of plaintext, but user data may be uploaded to the server in the form of ciphertext. Thus, there are mixtures of plaintext and ciphertext operations. Understanding what these stages are is important because it helps us understand how a single bit flip may be propagated by each stage. For instance, we can see that for plaintext addition PADD (Equation \ref{eq:padd}), a fault in the plaintext $p$ will only affect $c_{0}$, hence limiting the fault propagation. While in plaintext multiplication (Equation \ref{eq:pmul}), a fault in the plaintext $p$ propagates to both ciphertext components. Similar behavior can be deduced for ciphertext addition HADD. 


\paragraph{Fault resilience in ML}
When operations are stitched together to perform a larger computation, for example, a single layer in a neural network computation requires many additions, multiplications, and activations, error propagation may be extensive. Some bits in memory may not affect program correctness when corrupted, following an architectural vulnerability analysis~\cite{mukherjee2003systematic}. Other bits, while affecting program output, may not affect neural networks used in deep learning much. In many cases, the effects of errors result only in slight degradation in output accuracy~\cite{rakin2019bit}\footnote{This is not contrary to some works that find "certain" bits when flipped can cause significant errors~\cite{DNN_bitflip_2}, as random bitflips have a high probability of not affecting such bits.}. This leaves us the question of how neural network's natural fault tolerance is affected when running with FHE.

\paragraph{Fault detection timing}
Another implication for fault tolerance mitigation designs is that detecting the error earlier is better than later, before an error propagates broadly. This makes it more likely that computation can recover from the error. Furthermore, detecting an error late means that many computational resources have already been wasted. Given how resource-intensive FHE is, both spatially and temporally, early error detection is preferred to a late one.


\begin{table*}[tbp]
\centering
\small  
\caption{Comparison of Various Memory Fault Mitigation Approaches for FHE Computation}
\label{tab:fault-tolerance}
\renewcommand{\arraystretch}{1.1}  
\setlength{\tabcolsep}{3pt}  
\begin{tabular}{l*{4}{c}|cc|cc}  
\hline
& \textbf{ECC} & \textbf{TMR} & \textbf{Checksum} & \textbf{Output} & \textbf{DataSeal } & \textbf{zkOpenFHE} & 
\textbf{Modulo} & \textbf{OpenFHE} \\
& \textbf{(SECDED)} & & & \textbf{Check} & \textbf{\cite{santriaji2024dataseal}} & \textbf{\cite{knabenhans2023vfhe}} & 
\textbf{Check} & \textbf{DE-CKKS} \\
\hline
\makecell[ll]{Protection \\ Coverage} & 
\makecell[c]{Stored \\ Interm.} & 
\makecell[c]{Stored \\ Interm. \\ Ops} & 
Stored & 
\makecell[c]{Stored \\ Interm.} & 
\makecell[c]{Stored \\ Interm. \\ Ops} & 
\makecell[c]{Stored \\ Interm. \\ Ops} & 
\makecell[c]{Stored \\ Interm.} & 
\makecell[c]{Stored \\ Interm.} \\
\hline
\makecell[ll]{Detection Loc. \\ \& Timing} & 
\makecell[c]{Server \\ (runtime)} & 
\makecell[c]{Server \\ (runtime)} & 
\makecell[c]{Server \\ (runtime)} & 
\makecell[c]{Client \\ (post)} & 
\makecell[c]{Client \\ (post)} & 
\makecell[c]{Client \\ (post)} & 
\makecell[c]{Server \\ (runtime)} & 
\makecell[c]{Client \\ (post)} \\
\hline
FHE Scheme & 
\makecell[c]{CKKS \\ BGV} & 
\makecell[c]{CKKS \\ BGV} & 
\makecell[c]{CKKS \\ BGV} & 
\makecell[c]{CKKS \\ BGV} & 
\makecell[c]{BGV} & 
\makecell[c]{CKKS \\ BGV} & 
\makecell[c]{CKKS \\ BGV} & 
CKKS \\
\hline
Requirements & 
\makecell[c]{Hardware} & 
- & 
- & 
\makecell[c]{Output \\ expectations} & 
\makecell[c]{Transparent \\ circuit} & 
\makecell[c]{Transparent \\ circuit} & 
- & 
\makecell[c]{No  imaginary \\ part} \\
\hline

\end{tabular}
\end{table*}

\subsection{Ideal Characteristics of Mitigation Techniques} 

In this section, we will discuss characteristics of FHE computation and deduce several ideal characteristics for fault tolerance techniques for it. First, due to the spatial and temporal resource intensiveness of FHE computation, an ideal fault mitigation technique should have a low memory and low execution time overhead to avoid adding to the already-high memory pressure and execution time overheads. 

Second, to design an error detection, it is important to consider {\em coverage} (the percentage of errors that are detected) and {\em false positive rate} (the percentage of detected events that do not represent errors). Ideally, we would like a high coverage and a low false positive rate. 

Third, we need to consider \textbf{server-side verification capability}, which is the ability to detect a fault by the server without involving the client. This is particularly important in service-oriented deployments, such as Machine Learning as a Service (MLaaS), where service providers use proprietary models on client data \cite{wood2020homomorphic}.  Methods that enable servers to detect errors without client involvement minimize communication overhead and allow servers to proactively address faults, preventing the waste of significant computational resources. 

\subsection{Traditional Mitigation Techniques}

We explore potential bit-flip mitigation strategies for FHE systems by analyzing how these approaches address the unique challenges of FHE: large ciphertexts, computational intensity, and complex mathematical structures. We discuss four general approaches (TMR, ECC, checksum, and output checks) in this section, and then examine several FHE-specific possible approaches (modulo check and error estimation) in the next section.

Our qualitative analysis is summarized in Table~\ref{tab:fault-tolerance}. The table summarizes the key characteristics of various protection mechanisms, including their protection coverage, detection location and timing, and FHE scheme compatibility.


\subsubsection{\textbf{Error-Correcting Code (ECC) Memory}}
\label{subsubsec:ecc}
Error correcting code (ECC) memory \cite{hamming1950error} uses additional bits to detect and correct memory errors, based on the Hamming Code. The most commonly implemented ECC is Single Error Correction, Double Error Detection (SECDED) code that can correct single-bit errors and detect up to two-bit errors. ECC provides protection for both stored data and intermediate data by operating at the memory level, covering all data structures used during FHE computation. This server-side, runtime detection capability enables immediate fault handling without requiring client involvement.

A SECDED ECC adds 8 parity bits to each 64-bit data word, resulting in a 12.5\% space overhead, which is significant for FHE systems that already have massive ciphertexts. If implemented in hardware, ECC does not incur much additional execution time but is applied to the entire main memory. Due to its cost, ECC is typically only implemented in servers. On the other hand, if the hardware does not support it, ECC can be implemented in software, but in this case, ECC would incur substantial execution time overhead.

Stronger ECC codes that can handle multi-bit errors, e.g., 3EC2ED, 4EC3ED, etc., exist, but they come at much higher costs.~\cite{kim2007multi, wilkerson2010reducing}.


\subsubsection{\bf Triple Modular Redundancy (TMR)}
\label{subsubsec:tmr}

TMR~\cite{lyons1962use} offers strong protection by running three identical copies of a system and using majority voting to determine correct outputs. TMR is typically implemented by replicating entire computational sequences, providing comprehensive protection coverage, safeguarding stored data, intermediate data, and operations.

TMR operates server-side during runtime, enabling immediate fault detection and correction without client involvement. The approach is compatible with any FHE schemes. While TMR provides both detection and recovery capabilities, offering strong fault tolerance even against certain classes of multi-bit errors, it more than triples resource requirements in both computation and memory. This makes TMR prohibitively expensive for already resource-intensive FHE systems. Additionally, TMR can be vulnerable to correlated faults affecting multiple replicas simultaneously or attacks targeting the voting mechanism itself~\cite{aghaie2020impeccable}.

\subsubsection{\textbf{Checksum Methods}}
\label{subsubsec:checksum}

Checksum approaches compute a verification code for data at various processing stages to verify its integrity. These methods operate server-side during runtime and are compatible with both CKKS and BGV schemes, providing a lightweight detection mechanism with minimal overhead compared to redundancy-based approaches. However, checksums primarily protect stored data (e.g., machine learning model parameters) while the intermediate outputs are left unprotected. This limited protection coverage leaves intermediate computational results vulnerable, which is problematic for FHE systems with complex multi-stage operations. For instance, vector multiplication in the ciphertext space requires additional processing steps like relinearization and rescaling (See Equation \ref{eq:hmul}), creating more intermediate outputs that remain unprotected.


\subsubsection{\textbf{Output Check}}
\label{subsubsec:output_check}

The Output Check approach \cite{DNN_bitflip_2} works by monitoring the statistical patterns in computation outputs to detect anomalies that may indicate faults. Unlike the previous approaches, the Output Check strategy is done in the plaintext space, which means that the detection can only be done at the client side after computation. Even though this approach is not on the server side, Output Check is worth evaluating as it still fits with our fault model. Unlike several client-side integrity protection strategies discussed in~\cite{viand2023verifiable}, the Output Check strategy can work without knowledge of the server's computation circuit or parameters. 

However, there are other potential challenges with Output Check. Its effectiveness depends on the client having prior expectations or knowledge of what the output should look like. For instance, a straightforward strategy is to compare whether the current output falls into the statistical distribution of previous outputs (e.g., within mean ± standard deviation). But collecting the statistical distribution incurs computation overheads, and when the client does not have the previous output, defining the boundary will be more difficult. Another challenge for this approach is the possibility of false positives (i.e., false faults) when a legitimate output falls beyond the statistical norm of past outputs.

\subsubsection{\textbf{Strategies Requiring Public Input and Circuit}}
\label{subsubsec:public_circuit}

It is worth mentioning that several recent approaches that focus on verifying FHE computation require the knowledge of the computation circuit and input. For example, \textbf{(1) DataSeal} \cite{santriaji2024dataseal} applies Algorithm-Based Fault Tolerance principles to FHE, primarily designed for BGV, offering relatively efficient memory and computational overhead for matrix-based computation. \textbf{(2) zkOpenFHE} \cite{knabenhans2023vfhe} integrates Zero-Knowledge Proofs with OpenFHE to provide strong cryptographic guarantees. Both approaches offer robust protection and comprehensive coverage, but they fall outside our FHE computation model, as they require a public circuit where the client must know the exact computation performed by the server.






\subsection{FHE-inherent Bit-flip Mitigation Strategies}
\label{subsec:fhe_inherent_strategies}

Beyond conventional approaches, we examine several possible detection strategies that leverage the intrinsic properties of FHE schemes. These approaches have not been explored in prior work but merit consideration because they may naturally have low costs in terms of storage and computation.

\subsubsection{Modulo Check}
This lightweight verification mechanism exploits the modular arithmetic properties of FHE schemes. In both CKKS and BGV, all values in ciphertexts or plaintexts are therefore necessarily mathematically bounded by their respective moduli. In an RNS (Residue Number System) representation, as mentioned in \S\ref{sec:rns}, each value is represented across multiple ``slots'' with respect to different prime moduli $(q_0, q_1, \ldots, q_l)$. By design, all values in the $j$-th slot must remain strictly bounded by the modulus $q_j$.

A fault can cause values to exceed their modulus bounds. By implementing verification checks before computation that test whether any value exceeds its corresponding modulus, we can detect faults with zero memory overhead and negligible computation overhead. This approach operates directly on encrypted data, allowing server-side verification without compromising privacy.

\subsubsection{Dynamic Error Estimation in CKKS (DE-CKKS)}
\label{subsubsec:de-ckks}

OpenFHE implements the CKKS scheme that utilizes noise analysis techniques~\cite{costache2023precision, li2022securing}. This DE-CKKS scheme raises a warning during decryption when the noise is detected to be too large. As implemented after decryption, DE-CKKS requires secret keys and can only be implemented on the client side.  DE-CKKS utilizes the imaginary components of the ciphertext, reducing the CKKS capability to work only with the real number components. If our hypothesis proves correct, this method will have good robustness and protection coverage, while at the same time not incurring any memory or computation overhead. 


\section{Methodology}
\label{sec:methodology}

This section discusses the methodology of our experiments, starting from the general experimental setup such as workloads, fault injection method, and platforms (Section~\ref{subsec:general_setup}), and moving on to methodology for three research questions; RQ1 (Section~\ref{subsec:basic_operations}): how does a memory fault affect the result of individual FHE operation? RQ2 (Section~\ref{subsec:fhe_inference}): how does a memory fault affect the result of machine learning applications which are known to be naturally resilient to faults? and RQ3 (Section~\ref{subsec:fhe_mitigation}): how do various mitigation techniques compare in the context of FHE computation?   

We evaluate two popular FHE schemes, CKKS and BGV, with diverse parameter sets and scenarios. While other FHE schemes exist, such as TFHE~\cite{chillotti2020tfhe}, we focus on CKKS and BGV for several reasons: (1) They represent the two primary computation paradigms in FHE, supporting integer arithmetic (BGV) and real number arithmetic (CKKS), (2) TFHE uses fundamentally different computational primitives compared to the polynomial-ring operations that require different fault injection strategies and evaluation metrics. We leave TFHE for future work.

\subsection{General Experimental Setup}
\label{subsec:general_setup}

\subsubsection{FHE Libraries and Configurations}
For our experiments, we used OpenFHE (formerly PALISADE) version 1.3.1, a state-of-the-art open-source FHE library \cite{openfhe} with standard recommendations for 128-bit security. With these settings, OpenFHE automatically determines the ring dimension $N$ based on the targeted multiplicative depth $L$ and the size of the encrypted vector. We used \texttt{FIXEDMANUAL} CKKS scaling setting for basic operations experiments and \texttt{FIXEDAUTO} setting for machine learning experiments \cite{kim2022approximate}. The FHE parameters used are depicted in Table~\ref{tab:parameter_configurations}.

\begin{table}[tbp]
\centering
\small
\caption{FHE parameters used in basic operation experiments. Values in $\{\cdot\}$ indicate different values that we are going to evaluate; the first value is the default parameter.}
\label{tab:parameter_configurations}
\begin{tabular}{ll}
\hline
\textbf{Parameter} & \textbf{Values}  \\
\hline
Multiplicative Depth ($L$) & \{1, 3, 5\} \\
Ring Dimension ($N$) & \{$2^{13}$, $2^{14}$, $2^{15}$\}  \\
BGV Plaintext Modulus  & $65537$  \\
CKKS Scaling Factor  & $2^{50}$  \\
\hline
\end{tabular}
\end{table}

\subsubsection{Platform and Fault Injection}

All experiments were run on a workstation equipped with Intel(R) Xeon(R) W-2235 CPU @ 3.80GHz and 128 GB of DDR4 RAM.

We use a bit flip fault injection scheme based on prior works \cite{pytorchfi, ares}, specifically targeting transient bit flip faults that affect the ciphertext or plaintext stored in the memory hierarchy. This fault could affect not only the stored parameters, but also the input and intermediate output. We randomly flip a single bit uniformly distributed across the memory blocks, in line with \cite{rakin2019bit, ponader2021milr}.

To evaluate robustness beyond single-bit scenarios, we also inject burst bit flip faults where multiple bits (specifically 3 bits) are flipped simultaneously within the same 64-bit memory word. Recent empirical studies of DRAM faults show that real memory systems experience various multi-bit fault patterns affecting specific internal components~\cite{jung2023predicting}. Our burst injection serves as a stress test to evaluate system behavior under clustered fault scenarios that exceed the correction capabilities of common ECC schemes.

\subsection{Memory Fault on Basic FHE Operations (RQ1)}
\label{subsec:basic_operations}

\subsubsection{Tasks}
We evaluate all operations shown in Section \ref{heops}. All faults were injected into the ciphertext operand. While for PADD and PMULT, we also conducted experiments where the fault was injected into the plaintext operand; we refer to these as PADD-P and PMULT-P. Our experiments used vectors of 128 elements, with data ranges chosen to reflect typical machine learning applications. For CKKS, we used floating-point values in the range $[0..1]$, resembling normalized features in continuous data applications. For BGV, we used integers in the range $[0..255]$, corresponding to typical values in image processing tasks.

\subsubsection{Experiment Setup}
A single evaluation will only involve a single operation to be evaluated, injected with a single bit-flip fault. For instance, in the PADD-P evaluation, the workflow will be as follows: (1) we prepare one ciphertext $\mathbf{ct}$ and one plaintext vector $p$ (2) inject a fault in $p$ (3) perform PADD($\mathbf{ct},p$). To ensure statistical significance, we conducted 10,000 independent fault injection trials. 



\subsubsection{Evaluation}
To quantify the impact of faults, we measured the following key metrics:

\paragraph{Error Rate:} The percentage of fault injections that resulted in incorrect outputs across all trials.
    
    
\paragraph{Error Magnitude:} By using this metric, we can see how big the impact of a single bit flip is to the system's output. In addition, we can also characterize the result here for the Output Check scheme (Section \ref{subsubsec:output_check}). For operations other than HROT, we use the Median Absolute Error (MedAE):
    \begin{equation}
        \text{Error Magnitude} = \text{MedAE} = \text{median}(|X_{\text{orig}} - X_{\text{fault}}|)
    \end{equation}
    where $n$ denotes the number of evaluated elements, $ X_{\text{orig}}$ is the original output, and  $X_{\text{faulty}}$ is the faulty output. In HROT, we calculate the error magnitude using Hamming Distance:
    \begin{equation}
        \text{Error Magnitude} = \sum_{i=1}^{n} [X'_i \neq X_i]
    \end{equation}
    where $[\cdot]$ is the Iverson bracket.


\begin{figure}[tbp]
\centerline{\includegraphics[width=0.48\textwidth]{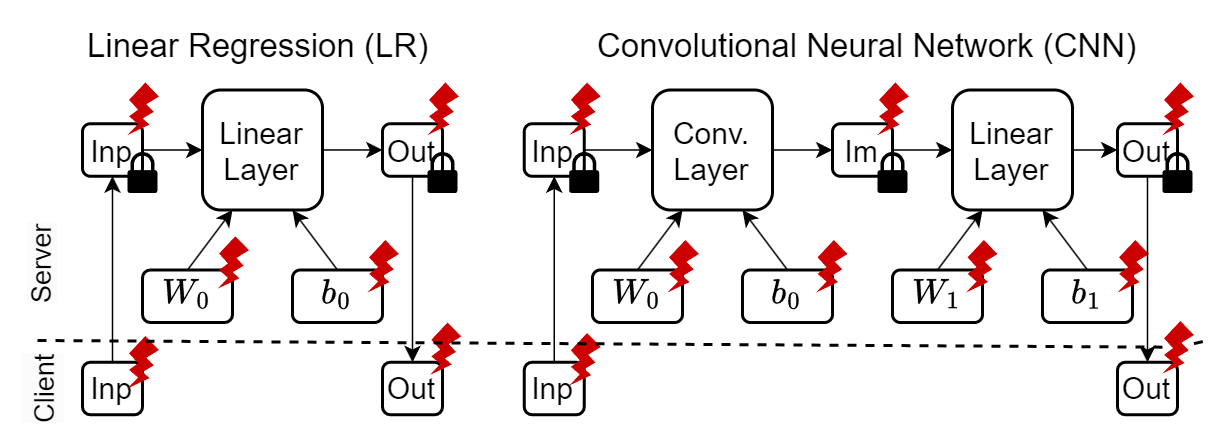}}
\caption{Illustration of fault location across machine learning workloads. Red bolts indicate a fault. $W$ and $b$ indicate the parameters of each layer, weight and bias, respectively.}
\label{figfault}
\end{figure}

\subsection{Memory Fault Effects on FHE-Protected Inference (RQ2)}
\label{subsec:fhe_inference}

\subsubsection{Tasks} 
Machine learning inference workloads are well known to be quite fault-tolerant, where they often exhibit a slight degradation in output accuracy when a single memory bitflip occurs. To test whether this is still the case with FHE computation, we chose two representative workloads that differ in their goals: value prediction vs. classification, derived from the publicly available challenge \cite{arakelov2024fherma}:

\textbf{California Housing Price Prediction.} We evaluated a simple Linear Regression (LR) model for the California Housing Price Prediction task \cite{pace1997sparse}. We use the publicly available training dataset version from FHERMA\footnote{https://fherma.io/challenges/}. We used a straightforward implementation of the \texttt{EvalInnerProduct} function from OpenFHE.

    
\textbf{CIFAR10 Image Classification.} We evaluated a privacy-preserving convolutional neural network (CNN) for encrypted CIFAR10 \cite{krizhevsky2009learning} image classification. We use a publicly available, trained model\footnote{https://github.com/UCF-ML-Research/FHERMA}. The model consists of one convolutional layer with a 3×3 kernel and 16 filters, followed by a square activation function, and one fully connected layer with 10 output neurons, adapted from \cite{rovida2024encrypted, juvekar2018gazelle}.


For both workloads, we followed the typical privacy preserving MLaaS paradigm where model parameters (weights and biases) are stored as plaintexts, while input features and prediction results remain encrypted throughout computation. This configuration represents a practical scenario where a service provider offers prediction capabilities without accessing the raw user data. Both workloads were implemented using the CKKS scheme, which is well-suited for approximate floating-point computations required in machine learning tasks. The FHE parameters are shown in Table~\ref{tab:inference_parameters}. Due to the long computation required for private inference, we only use 500 stratified samples from each dataset.

\begin{table}[tbp]
\centering
\small
\caption{Parameter configurations for inference workloads}
\label{tab:inference_parameters}
\begin{tabular}{lll}
\hline
\textbf{Parameter} & \textbf{LR} & \textbf{CNN} \\
\hline
Multiplicative Depth ($L$) & 4 & 6 \\
Ring Dimension ($N$) & $2^{14}$    & $2^{15}$ \\
CKKS Scaling Factor   & $2^{50}$  & $2^{50}$ \\
\hline
\end{tabular}
\end{table}

\subsubsection{Experiment Setup} 

We conducted a comprehensive fault location analysis as illustrated in Figure~\ref{figfault}. We systematically injected faults in 15 different locations of the computation model shown in the figure, including the input ciphertext (Inp), model parameters ($W$ and $b$), and intermediate results at the server side (Im), and at the encoded but not encrypted plaintext at the client side (shown as Inp without a lock icon). We distinguish the injection into the Weight parameters ($W$) that apply multiplication and addition, and the Bias ($b$) that only involves the addition.

\subsubsection{Evaluation Metrics}
To evaluate fault impact on these workloads, we used task-specific metrics. For linear regression, we used the median absolute error (\textbf{MedAE}) to quantify prediction quality, measuring how well the predictions approximate the actual data. For the neural network, we measured classification \textbf{Accuracy}.

\begin{figure*}[tbp]
\centerline{\includegraphics[width=0.94\textwidth]{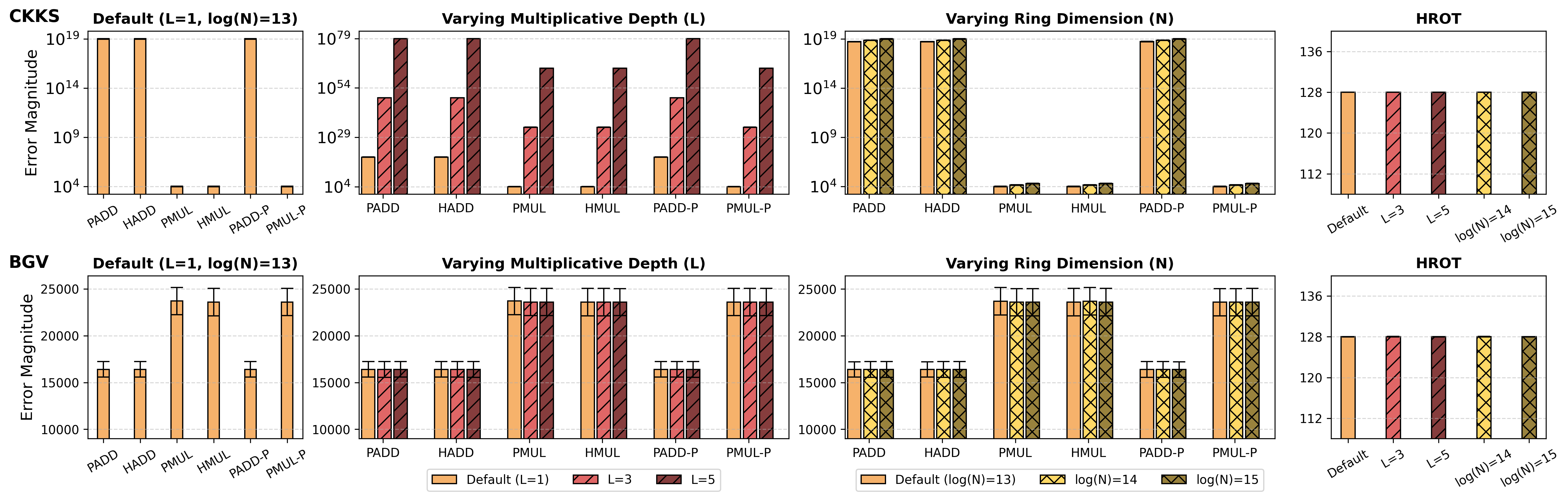}}
\caption{Average error magnitude of single-bit flips on FHE operations with varying security parameters, first row for CKKS, second row for BGV. The first column shows the error magnitude in the default parameter setting. The second and third columns show error magnitude when the multiplicative depth ($L$) and ring dimension ($N$) are varied, respectively. The fourth column shows the error magnitude for the HROT operation. The vertical line on top of a bar indicates the standard deviation}
\label{fig:impact}
\end{figure*}


\subsection{Evaluated Mitigation Techniques}
\label{subsec:fhe_mitigation}

We evaluate potential fault detection and mitigation approaches for the FHE systems mentioned earlier. We evaluated on ciphertext multiplication (HMUL) with a single bit flip. For {\bf Triple Modular Redundancy (TMR)}, we implemented it at both the operational level (replicating the entire computational sequences) and the data level (replicating data ciphertexts and plaintexts). We use the voting scheme to verify the correctness. We assume the voting module incurs negligible overhead compared to the redundant data and computation. We implemented {\bf Error-Correcting Code (ECC)} in the software using the common single-error correction and double-error detection Hamming code (72, 64) scheme. The {\bf Checksum Methods (CS)} used the CRC32 algorithm to verify data prior to computation by comparing the stored golden value. For the {\bf Output Check (OC)}, we obtained a golden distribution (average and standard deviation) using 50 data samples. If the system output falls beyond the average $\pm$ standard deviation, we assume that it is an incorrect computation. We implemented the {\bf Modulo Check (MC)} by checking whether each value in the ciphertext exceeded its modulus before and after computation. Due to the mathematical use of polynomial modulus, any value exceeding the modulus is guaranteed to be erroneous. Finally, for the {\bf DE-CKKS}, we use noise detection feature from OpenFHE ~\cite{costache2023precision}. 

To address the limitations of individual techniques, particularly when dealing with burst bit flip scenarios, we also evaluated two combined approaches. The first one, \textbf{ECC+Modulo Check (EM)}, augments  ECC's inability to detect faults involving three or more bits with the modulo check serves as a backup detection mechanism for multi-bit faults that exceed ECC's correction capabilities. The second, \textbf{ECC+Modulo Check+Checksum (EMC)} further adds a checksum to detect faults in stored data, such as plaintext weights, allowing the modulo check to focus specifically on intermediate computational results that bypass both ECC capabilities and stored data protection.

In this evaluation, we measured three key metrics. The \textbf{Detection Rate} reflects the percentage of injected faults successfully identified by each approach to measure the effectiveness of each detection strategy. The  \textbf{Computation Overhead} measures the additional computation time required for detection, relative to the case with no mitigation. The \textbf{Memory Overhead} measures the extra memory requirements for each approach, relative to the case with no mitigation. Computation and memory overheads are particularly important for FHE systems due to their much longer execution time and larger ciphertext sizes.
    
    
    


\section{Results}
\label{sec:results}
\subsection{Results for Basic FHE Operations (RQ1)}

Our first set of experiments injects a single bitflip for each FHE operation, and we observe the effects on basic FHE operations. The result is shown in Figure~\ref{fig:impact}. In all of our experiments, memory bitflip always leads to incorrect results for both CKKS and BGV. Thus, virtually every bit is required for an architecturally correct execution (ACE) of an FHE operation.

For CKKS, the error magnitudes are massive, up to $10^{19}$ for addition operations. In contrast, multiplication operations have relatively smaller errors but are still significant,  around $10^{4}$. This low error magnitude in multiplication operations can be attributed to the rescaling process applied after multiplication in CKKS (Section~\ref{heops}), which scales down the noise and leads to a smaller error magnitude. For BGV, the error magnitudes are more consistent and significantly lower than for CKKS. This stability stems from BGV's use of a plaintext modulus $t$, which constrains the numeric range of potential errors after decryption. However, even for BGV, the errors are not small by any means.




\begin{observation}
A single-bit flip in FHE computation consistently leads to significant errors in each individual FHE operation, for both CKKS and BGV. The error magnitude on all operations is also large, the smallest one has median absolute error of $\approx10^4$ and can be up to $\approx10^{79}$
\end{observation}





\begin{figure*}[tbp]
\centerline{\includegraphics[width=\textwidth]{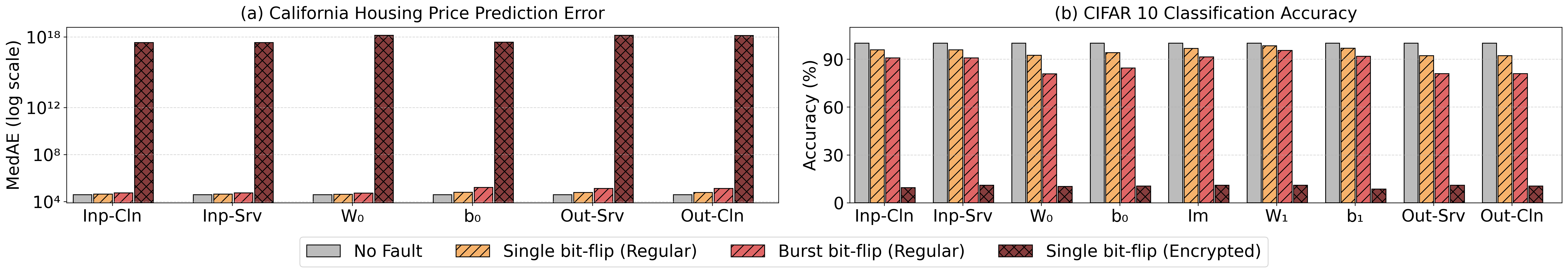}}
\caption{Impact of memory faults on the MedAE of the House Pricing Prediction task (Left) and on the accuracy of the CIFAR10 image classification task across different fault positions (Right). Fault position naming follows Figure \ref{figfault}, with *-Cln means plaintext in the client and *-Srv means ciphertext in the server.}
\label{fig:nn_linreg}
\end{figure*}

\subsubsection{Varying parameters} 

The result of varying multiplicative depth ($L$) and ring dimension ($N$) can be seen in the second and third columns of Figure~\ref{fig:impact}.

For CKKS, increasing the multiplicative depth dramatically amplifies error magnitudes, growing from $10^{19}$ at depth 1 to $10^{79}$ at depth 5 for addition operations. This growth indicates that deeper computation circuits become more complex. This is problematic because a larger depth allows more operations to be performed, but with memory faults, the errors get magnified much more. In contrast, BGV maintains relatively stable error magnitudes across different parameters, suggesting inherent differences in how these schemes propagate errors. Ring dimension variations show a more linear relationship with error magnitude. For CKKS operations, the average error magnitude approximately doubles as the ring dimension increases from $2^{13}$ to $2^{15}$.


\subsection{Memory Fault in FHE-Protected Inference (RQ2)}

For this research question, we examine how memory faults affect FHE applications. Figure~\ref{fig:nn_linreg} (Left) shows the impact on median absolute error (MedAE) of single-bit flips at different layers of the house pricing prediction task, while Figure~\ref{fig:nn_linreg} (Right) shows the impact on accuracy in the image classification model. For both figures, the first bars in each group represent no-fault and non-FHE situation. The second and third bars represent single bitflip and a burst of bitflips situations affecting non-FHE computation. The fourth/final bars represent single bitflips on FHE computation.

Figure~\ref{fig:nn_linreg} (Left) shows that the predicted housing prices stay pretty close in systems without FHE computation (MedAE in the tens of thousands of dollars), but the errors become so huge even with single bitflips with FHE (MedAE nearly a trillion dollars), making the model unusable. 
Figure~\ref{fig:nn_linreg} (Right) shows a CNN for classification that shows non-FHE computation quite resistant to memory bitflips (about 90\% accuracy with single bitflips and close to 80\% for burst bitflips). These results are consistent with prior studies on NN robustness against memory faults~\cite{rakin2019bit,DNN_bitflip_2, agarwal2023resilience}. In contrast, for FHE computation, even single bitflips reduce classification accuracy to random levels (approximately 10\% on the CIFAR-10 subset). These accuracy drops occur at all fault locations and timings in the FHE pipeline, either in the plaintext before encryption on the client side (Inp-Cln), in the intermediate result (Im), server parameters ($W$ and $b$), or the output. The consistent degradation performance demonstrates that FHE-protected neural networks completely lost the natural fault tolerance seen in plaintext neural networks. This difference stems from the complex mathematical structures in FHE that amplify and propagate errors throughout the computation.



\begin{observation}
FHE-protected machine learning models lack the natural fault tolerance seen in plaintext models. A single bit flip in the pipeline, either on the client or server side, reduces classification performance to random levels and produces high regression errors.
\end{observation}

\begin{figure*}[htbp]
\centerline{\includegraphics[width=0.87\textwidth]{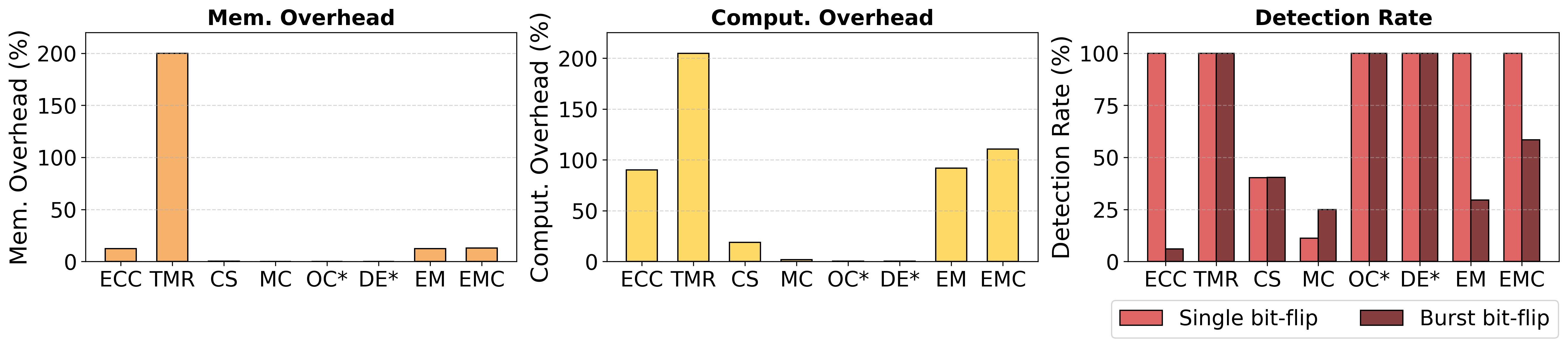}}
\caption{Comparison of Single Bit Flip Detection Methods for FHE Systems measured on single FHE operation. The X-axis is the short name of the evaluated methods mentioned in Section \ref{subsec:fhe_mitigation}. Asterix (*) in OC and DE indicates they are client-side protection that requires client decryption.}
\label{fig:fault-detection-comparison}
\end{figure*}

\subsection{Fault Propagation in FHE Systems}
As discussed in our earlier analysis (Section~\ref{sec:bkground}), with the polynomial representation used by modern FHE systems, such as Double CRT, element-wise operations like PADD and HADD exhibit relatively contained fault propagation, where a single-bit flip typically remains isolated to a single ciphertext component. 
In contrast, multiplication operations demonstrate more complex propagation patterns. In PMUL (Equation \ref{eq:pmul}), a fault in the plaintext $p$ affects both ciphertext components. HMUL and HROT operations exhibit substantially more severe fault propagation patterns due to complex transformations. These operations involve multiple stages of computation, including data transformation using Number Theoretic Transform (NTT). 
We suspected that it was NTT that amplifies the impact of even a single-bit error.

To test this, we checked the resulting polynomials after NTT propagates a single bitflip and found that 100\% of the ciphertext polynomial coefficients contain errors. This observation also gives an insight other vulnerable steps in FHE: since NTT is used not only in operations such as HMUL and HROT, but its inverse is also used in the decoding stage, a mandatory step after decryption in Double-CRT representation, then the decoding stage is also vulnerable to diffusing a single bitflip into massive bit errors.


Fault propagation in the decoding stage strengthens the motivation to contain the fault on the server side early, before passing it to the client for decryption and decoding. However, as mentioned earlier, NTT is not only used in the decoding but also in other operations such as HMUL and HROT. In the House Pricing Prediction task, a linear regression computation $XW + b$ does not seem to involve a complex operation, as $W$ and $b$ are plaintexts. However, OpenFHE's \texttt{EvalInnerProduct} operation, which is used for computing $XW$, involves HROT to sum up the values after multiplication. Therefore, it will propagate the fault to the entire ciphertext. This demonstrates that containing faults in FHE systems requires a good understanding of the FHE computation performed by the system. This is particularly concerning for more complex FHE systems that chain multiple operations together, making traditional error detection approaches insufficient for comprehensive protection.

\begin{observation}
NTT/Inverse NTT propagates a single-bit flip throughout the entire data structure. Since NTT is used in operations like HMUL and HROT, any computation involving them are vulnerable to massive errors. Furthermore, since inverse NTT is used for  decoding to transform the polynomial representation into the original output, even a single bitflip in FHE systems will corrupt the entire output after decoding. This also means that the client side also needs to have good error protection capability. 
\end{observation}

\subsection{Evaluating potential fault detection (RQ3)}

To avoid catastrophic failures of FHE computation from being undetected, we need robust fault detection mechanisms for reliable FHE deployment. We evaluated several potential detection strategies, summarized in Figure~\ref{fig:fault-detection-comparison}. 


\subsubsection{\textbf{Redundancy approaches}} Traditional redundancy techniques face significant hurdles. ECC memory, while offering detection and correction for single-bit errors and detection (but not correction) for double-bit errors, imposes a non-trivial memory overhead (12.5\%). More critically, ECC memory completely fails to detect errors involving three or more bits (6.0\% detection rate), creating a fundamental risk for FHE systems where, as our experiments demonstrate, even a single bit flip causes 100\% output corruption. ECC memory hardware has negligible time overhead, but the requirement for specific ECC hardware potentially limits the choice of hardware accelerators available for FHE systems. On the other hand, software ECC incurs a significant overhead (90.1\%). Triple Modular Redundancy (TMR) provides more comprehensive protection: 100\% detection rate in both fault scenarios with recovery, but its more than 200\% time and memory overheads are fundamentally incompatible with the resource-intensive nature of FHE.

\subsubsection{\textbf{Low-cost approach}} Checksum and modulo checking offer significantly reduced overheads (18.8\% and 1.5\% time overhead, respectively) but sacrifice detection capability (40.3\% and 11.2\% detection rates). This limited detection rate for checksums stems from their narrow protection coverage, as they only safeguard stored data while leaving intermediate computational results vulnerable. Modulo checking performs even worse (11.2\%) because its detection mechanism only triggers when values exceed the moduli, which is a relatively rare occurrence that misses most corruption patterns in FHE's complex polynomial structures.


\subsubsection{\textbf{Client-side approach}}


The Output Check strategy, which seems simple, is effective for detecting faults in FHE systems. Different from regular systems \cite{DNN_bitflip_2}, faults in FHE systems significantly shift the output during the decoding process, making fault detection by output analysis quite robust. In addition, because the checking is done in the plaintext, which is smaller than the ciphertext, the checking process is much faster than the Modulo Check strategy ($< 1\%$ time overhead). DE-CKKS also shows a similar result; by sacrificing the imaginary part of the data, it can achieve robust fault detection with a very low computation overhead. This low cost of DE-CKKS is achieved as they leverage the inherent property of the CKKS algorithm. Both strategies also have fewer requirements compared to prior works that work on the client side \cite{santriaji2024dataseal, viand2023verifiable, chatel2022verifiable} by not requiring transparent server input and circuit. 


\subsubsection{Combined protection strategy} 
To address ECC's critical weakness in burst fault scenarios, we evaluated two combined approaches. The ECC + Modulo Check (EM) strategy addresses ECC's limitations by using modulo checking as a backup detection mechanism for multi-bit faults that exceed ECC's correction capabilities. However, EM achieves only 29.6\% burst fault detection. The three-layer ECC + Modulo Check + Checksum (EMC) approach improves burst detection to 58.5\% by adding checksum verification for stored data protection. While EMC represents the most effective server-side approach for burst fault detection, it still leaves 41.5\% of burst faults undetected.

\begin{observation}
None of the evaluated mechanisms provides a satisfactory balance of a high detection rate, low overhead (both in terms of time and memory), and server-side operability for general FHE schemes. The methods with high detection rates impose significant overhead that is incompatible with FHE's scale (TMR, ECC) or require client involvement, detecting errors post-computation (Output Check, DE-CKKS). Conversely, low-overhead server-side methods (Checksum, Modulo Check) are demonstrably ineffective due to FHE's nature. Combined protections are likely needed, but more research is needed to boost the detection rates. 
\end{observation}



\section{Conclusion}
\label{sec:conclusion}
We demonstrated how a single bit flip leads to massive errors in the FHE systems. At the application level, even typically fault-tolerant machine learning applications lose their natural fault tolerance. We explored various memory fault mitigation techniques, both traditional (ECC, TMR, checksum) and FHE-specific (output check, noise estimation). We found that each has a substantial limitation that cannot simultaneously achieve low overheads, high detection rates, and universality across FHE schemes. These findings highlight a critical gap in FHE system design that must be addressed in the future, as reliability challenges become increasingly urgent with the diffusion of real-world deployments of FHE.                                                                                                                                                                                                                                                                                                                                                                                                                                                                                                                                                                                                                                                                                                                                                                                                                                                                                                                                                                                                                                                                                                                                                                                                                                                                                                                                                                                                                                                                                                                                                                                                                                                                                                                                                                                                                                                                                                                                                                        









\bibliographystyle{ACM-Reference-Format}
\bibliography{references}










\end{document}